\documentstyle[prd,eqsecnum,aps,amsfonts]{revtex}

\begin{document}
\draft
\title{Existence of maximal hypersurfaces in some
spherically symmetric spacetimes}
\author{Gregory A. Burnett}
\address{Department of Physics, University of Florida,
Gainesville, Florida\ \ 32611, USA}
\author{Alan D. Rendall}
\address{Institut des Hautes Etudes Scientifiques, 35 Route de
Chartres, 91440 Bures sur Yvette, France}
\date{1 August 1995}
\twocolumn[
\maketitle
\widetext
\begin{abstract}
We prove that the maximal development of any spherically symmetric
spacetime with collisionless matter (obeying the Vlasov equation) or a
massless scalar field (obeying the massless wave equation) and
possessing a constant mean curvature $S^1 \times S^2$ Cauchy surface
also contains a maximal Cauchy surface.  Combining this with previous
results establishes that the spacetime can be foliated by constant
mean curvature Cauchy surfaces with the mean curvature taking on all
real values, thereby showing that these spacetimes satisfy the
closed-universe recollapse conjecture.  A key element of the proof, of
interest in itself, is a bound for the volume of any Cauchy surface
$\Sigma$ in any spacetime satisfying the timelike convergence
condition in terms of the volume and mean curvature of a fixed Cauchy
surface $\Sigma_0$ and the maximal distance between $\Sigma$ and
$\Sigma_0$.  In particular, this shows that any globally hyperbolic
spacetime having a finite lifetime and obeying the
timelike-convergence condition cannot attain an arbitrarily large
spatial volume.
\end{abstract}
]
\narrowtext

\section{Introduction} \label{sec:intro}

Given an initial data set for the gravitational field and any matter
fields present, what can be said of the spacetime evolved from this
initial data?

In the asymptotically flat case, one would like to know such things as
how much gravitational energy is radiated to null infinity, the final
asymptotic state of the system, whether black holes are formed, the
nature of any singularities produced, and whether cosmic censorship is
violated.  For example, it is known that the maximal development of
sufficiently weak vacuum initial data is an asymptotically flat
spacetime that is free of singularities and black holes \cite{CK93}.
In this case the gravitational waves are so weak that they cannot
coalesce into a black hole; instead they scatter to infinity.  Further
it is known that an initial data set containing a future trapped
surface or a future trapped region must be singular, provided the
null-convergence condition holds \cite{HawkingEllis73,Wald84}.  In
these cases, the gravitational field is already sufficiently strong
that collapse is inevitable.

In the cosmological case (spacetimes with compact Cauchy surfaces),
the questions one asks are a bit different as one expects these
spacetimes to be quite singular.  In fact, it is known that spacetimes
with compact Cauchy surfaces are singular, provided a genericity
condition and the timelike-convergence condition hold
\cite{HawkingEllis73,Wald84}.  So, here one would like to know such
things as the nature of the singularities, if the spacetime has a
finite lifetime (in the sense that there is a global upper bound on
the lengths of all causal curves therein), whether it expands to a
maximal hypersurface and then recollapses or is always expanding
(contracting), and whether cosmic censorship is violated.  For
example, it is known that if the initial data surface is contracting
to the future (past), then any development satisfying the
timelike-convergence condition must end within a finite time to the
future (past) \cite{HawkingEllis73,Wald84}.  Can more be said about
the behavior of the cosmological spacetimes?

The closed-universe recollapse conjecture asserts that the spacetime
associated with the maximal development of an initial data set with
compact initial data surface expands from an initial singularity to a
maximal hypersurface and then recollapses to a final singularity (all
within a finite time), provided that the spatial topology does not
obstruct the existence of a maximal Cauchy surface (e.g., $S^3$ or
$S^1 \times S^2$) and provided the matter satisfies certain energy and
regularity conditions \cite{MarsdenTipler80,BarrowTipler85,BGT86}.  It
has also been conjectured that such spacetimes admit a unique
foliation by constant mean curvature (CMC) Cauchy surfaces with the
mean curvatures taking on all real values.  (See, e.g., conjecture~2.3
of \cite{EardleySmarr79} and the weaker conjecture~C2 of
\cite{EardleyMoncrief81}.)\ \ Just what energy conditions the matter
must satisfy is an open problem. However, in the study of the weak
form of this conjecture (which merely asserts that the spacetime has a
finite lifetime), the dominant energy and non-negative pressures
conditions together have proven sufficient for the cases studied
\cite{Burnett95,Burnett91}.  More subtle is the problem of what regularity
conditions the matter needs to satisfy.  The difficulty here is that
the maximal development of an Einstein-matter initial data set may not
contain a maximal hypersurface because of the development of a
singularity in the matter fields, such as a shell-crossing singularity
in a dust-filled spacetime, before the spacetime has a chance to
develop a maximal hypersurface.  While not for certain, it is thought
that those matter fields that do not develop singularities when
evolved in fixed smooth background spacetimes will not lead to the
obstruction of a maximal hypersurface.

Here, we study the maximal development of spherically symmetric
constant mean curvature initial data sets with $S^1 \times S^2$ Cauchy
surfaces and matter consisting of either collisionless particles of
unit mass (whose evolution is described by the Vlasov equation) or a
massless scalar field (whose evolution is described by the massless
wave equation).  It has already been established that if the mean
curvature is zero on the initial data surface, i.e., it is a maximal
hypersurface, then its maximal evolution admits a foliation by CMC
Cauchy surfaces with the mean curvature taking on all real values
\cite{Rendall95}.  Further, it is known that if the mean curvature is
negative (positive) then the initial data can be evolved at least to
the extent that the spacetime can be foliated by CMC spatial
hypersurfaces taking on all negative (positive) values
\cite{Rendall95}.  Left unresolved was whether the maximal evolution
in the latter two cases actually contains a maximal spatial
hypersurface and, hence, can be foliated by CMC hypersurfaces taking
on all real values.  The nonexistence of a maximal spatial
hypersurface would be reasonable if such spacetimes could expand
(contract) indefinitely, however, it is known that these spacetimes
have finite lifetimes \cite{Burnett95,Burnett91}.  Therefore, it would
seem that their maximal development should contain a maximal
Cauchy surface.  We show that it does.

{\it Theorem 1.}  The maximal development of any spherically symmetric
spacetime with collisionless matter (obeying the Vlasov equation) or a
massless scalar field (obeying the massless wave equation) that
possesses a CMC $S^1 \times S^2$ Cauchy surface $\Sigma$ admits a
unique foliation by CMC Cauchy surfaces with the mean curvature taking
on all real values.  In particular, it contains a maximal Cauchy
surface and its singularities are crushing singularities.

By the maximal development of a globally hyperbolic spacetime, we mean
the maximal development of an initial data set induced on a Cauchy
surface in the spacetime.  This is well-defined as the maximal
developments associated with any two Cauchy surfaces are necessarily
isometric \cite{CBGeroch69}.  Further, recall that a spacetime with
compact Cauchy surfaces is said to have a future (past) crushing
singularity if the spacetime can be foliated by Cauchy surfaces such
that the mean curvature of these surfaces tends to infinity (negative
infinity) uniformly to the future (past).  That the future and past
singularities associated with the spacetimes of theorem~1 are crushing
is then a simple consequence of the existence of a CMC foliation
taking on all real values.

As a consequence of theorem~1, the maximal development of the
spacetimes studied is rather simple.  They expand from an initial
crushing singularity to a maximal hypersurface and then recollapse to
a final crushing singularity---all in a finite time.  That is, they
satisfy the closed-universe recollapse conjecture in its strongest
sense as well as the closed-universe foliation conjecture.

While the maximal development of the spacetimes in theorem~1 is about
as complete as one could expect given the existence of a complete CMC
foliation, these spacetimes may still be extendible (though there is
no globally hyperbolic extension).  In other words, theorem~1 does not
eliminate the possibility that these spacetimes violate cosmic
censorship.  In fact, cosmic censorship is violated in the vacuum
case.  This is easily seen by realizing that the maximal development
in this case is either of the of the two regions where $r < 2M$ of an
extended Schwarzschild spacetime of mass $M$ ($r$ is the areal
radius), modified by identifications so that the Cauchy surface
topology is $S^1 \times S^2$.  Although the ``singularity''
corresponding to $r \to 2M$ is a crushing singularity, this is
actually a Cauchy horizon.  Is this vacuum case exceptional?  It is
worth noting that if a crushing singularity corresponds to $r \to 0$,
then the singularity must in fact be a curvature singularity.  This
follows easily from the fact that $R_{abcd} R^{abcd} \ge (4m/r^3)^2$,
for any spherically symmetric spacetime satisfying the
null-convergence condition, and the fact that the mass function $m$ is
bounded away from zero by a positive constant in our case
\cite{Burnett91}.  If we could show that $r$ must go to zero
(uniformly) at the extremes of our foliation, then the spacetime would
indeed be inextendible, thereby satisfying the cosmic censorship
hypothesis.  Establishing such a result appears to be difficult and
the vacuum case shows that such a result will not always hold (though
this case may be exceptional).  Using a different approach, Rein has
shown that for an open set of initial data, there is a crushing
singularity in which $r \to 0$ uniformly, and which, therefore, is a
curvature singularity \cite{Rein}.  While this is encouraging, the
extent to which the spacetimes of theorem~1 satisfy cosmic censorship
remains to be seen.

The proof of theorem~1 involves a combination of three ideas.  First,
it is known that spherically symmetric spacetimes with $S^1 \times
S^2$ or $S^3$ Cauchy surfaces and satisfying the dominant energy and
non-negative pressures (or merely ``radial'' non-negative pressure)
conditions have finite lifetimes \cite{Burnett95,Burnett91}.  Second,
using a general theorem (which is independent of symmetry assumptions)
established in Sec.~\ref{sec:volume}, it follows that the spatial
volumes of Cauchy surfaces in the spacetime are bounded above, which
allows us to bound various fields describing the spacetime geometry.
Third, introducing a new time function to avoid the problems
associated with ``degenerate'' maximal hypersurfaces (i.e., surfaces
where the mean curvature cannot be used as a good coordinate), the
theorem then follows using the methods developed in \cite{Rendall95}.
Furthermore, it is worth noting that our method uses only a few
properties of the matter fields themselves.  Namely, we use the fact
that they satisfy the dominant energy and ``radial'' non-negative
pressures conditions and, roughly speaking, the fact that the matter
fields are nonsingular as long as the spacetime metric is nonsingular.
This latter property has not been given a precise formulation, as it
seems difficult to do so, and serves merely as a heuristic
principle---the arguments for collisionless matter and the massless
scalar field in \cite{Rendall95} providing an example of what it means
in practice.

In theorem~1 we have restricted ourselves to spacetimes with $S^1
\times S^2$ Cauchy surfaces and have not considered similar spacetimes
with $S^3$ Cauchy surfaces.  The problem with the $S^3$ case is that
there exist two timelike curves on which the symmetry orbits
degenerate to points.  When we then pass to the quotient of our
spacetime by the symmetry group, the field equations on the quotient
spacetime are singular on boundary points corresponding to the
degenerate orbits.  Experience has shown that this degeneracy can have
nontrivial consequences on the evolution of the spacetime.  For
example, in the study of the spherically symmetric asymptotically flat
solutions of the Einstein-Vlasov equations, it has been shown that if
a solution of these equations develops a singularity, then the first
singularity (as measured in a particular time coordinate) is at the
center \cite{RRS}.  However, currently it is not known how to decide
when a central singularity must occur.  In the case of asymptotically
flat spherically symmetric solutions of the Einstein equations coupled
to a massless scalar field, Christodoulou has shown that naked
singularities do form in the center of symmetry for certain initial
data (and that they can form nowhere else) \cite{Christodoulou}.  Note
that the degeneracy of the orbits in these spacetimes is of the same
type that occurs in the spherically symmetric spacetimes with $S^3$
Cauchy surfaces.  Similar problems occur in the study of the vacuum
spacetimes with $U(1) \times U(1)$ symmetry and having $S^3$ or $S^1
\times S^2$ Cauchy surfaces.  Here the dimension of the orbits is
non-constant and, consequently, this case is much harder to analyze
than the $T^3$ case, which has orbits of constant dimension
\cite{Chrusciel}.  The spherically symmetric spacetimes with $S^1
\times S^2$ Cauchy surfaces, having no degenerate orbits, avoid these
complications.

It would, of course, be preferable to strengthen theorem~1 by removing
the requirement that there exist a CMC Cauchy surface in the
spacetime.  While such a result seems plausible, the methods currently
used are not adequate to cover this more general case.  Strengthening
our results in this direction is a subject for future research.

Our conventions are those of \cite{Wald84}, with the notable exception
that trace $H$ of the extrinsic curvature $K_{ab}$ of a spatial
hypersurface measures the {\em convergence} of the hypersurface to the
future.  Thus, surfaces with negative $H$ are expanding to the future,
while those with positive $H$ are contracting to the future.

\section{Proof of theorem~1} \label{sec:proof}

Fix a spacetime $(M,g)$ satisfying the conditions of theorem~1.  Both
classes of spacetimes considered here (the Einstein-Vlasov and
massless scalar field spacetimes) satisfy the dominant energy
condition (the Einstein tensor $G_{ab}$ satisfies $G_{ab}v^aw^b\ge 0$
for all future-directed timelike vectors $v^a$ and $w^b$) as well as
the timelike-convergence condition (the Ricci tensor satisfies
$R_{ab}t^a t^b \ge 0$ for all timelike $t^a$).  While the
Einstein-Vlasov spacetimes also satisfy the non-negative pressures
condition ($G_{ab}x^ax^b \ge 0$ for all spacelike $x^a$), in general
the massless scalar field spacetimes do not.  However, they do satisfy
the weaker ``radial'' non-negative pressures condition ($G_{ab}x^a x^b
\ge 0$ for all spatial vectors $x^a$ perpendicular to the spheres of
symmetry).  It was shown in \cite{Burnett95,Burnett91} that the
spherically symmetric spacetimes with $S^3$ or $S^1 \times S^2$ Cauchy
surfaces satisfying the dominant energy and the non-negative pressures
conditions (or merely the ``radial'' non-negative pressures condition)
have a finite lifetime, i.e., the supremum of the lengths of all causal
curves is finite.  Therefore, our spacetime $(M,g)$ has a finite
lifetime.  It then follows immediately from lemma~2 (established in
Sec.~\ref{sec:volume}) that the volumes of all spatial Cauchy surfaces
in $(M,g)$ are bounded above.

Denote the mean curvature of the Cauchy surface $\Sigma$ by $t_0$.
This initial data surface must be spherically symmetric.  In the case
$t_0 \neq 0$, this follows from the uniqueness theorem for such
hypersurfaces (see, e.g., theorem~1 of \cite{MarsdenTipler80}) since
if a rotation did not leave $\Sigma$ invariant, we would have a
distinct CMC Cauchy surface with identical (nonzero) constant mean
curvature.  The case where $t_0 = 0$ then follows from the fact that
there is a neighborhood $N$ of $\Sigma$ in $M$ such that $N$ can be
foliated by CMC hypersurfaces, each having a different CMC, and the
fact that those with non-zero CMC must be spherically symmetric.  As
the theorem has already been proven in the case where $t_0 = 0$
($\Sigma$ is a maximal hypersurface) \cite{Rendall95}, we shall take
$t_0$ to be negative ($\Sigma$ is expanding to the future).  The case
where the mean curvature is initially positive follows by a
time-reversed argument.  As was shown in \cite{Rendall95}, in a
neighborhood of the hypersurface $\Sigma$, the spacetime can be
foliated by CMC Cauchy surfaces.  Define the scalar field $t$ at any
point to be the value of the mean curvature of the CMC hypersurface
passing through that point, i.e., so level surfaces of $t$ are CMC
hypersurfaces and, in particular, the surface $t=t_0$ is $\Sigma$.  A
further scalar field $x$ can then be introduced so that the spacetime
metric $g$ is given by
\begin{equation} \label{metric}
g = -\alpha^2 {\rm d} t^2 + A^2[({\rm d} x+\beta {\rm d} t)^2 + a^2
\Omega],
\end{equation}
where $\Omega$ is the natural unit-metric associated with the spheres
of symmetry.  The functions $\alpha$, $\beta$, and $A$ depend only on
$t$ and $x$ (being spherically symmetric) and are periodic in $x$.
The function $a$ depends only on $t$.  The fields can be chosen so
that $\int \beta(t,x)\; {\rm d} x = 0$ for each $t$, where the
integral is taken over one period of a surface of constant $t$.

It was shown in \cite{Rendall95} that the initial data induced on
$\Sigma$ can be evolved so that $t$ covers the interval $(-\infty, 0)$
and that, if it can be evolved to the closed interval $(-\infty,0]$,
i.e., a maximal hypersurface is attained, the spacetime can be
extended and foliated by CMC spatial hypersurfaces taking on all real
values.  Therefore, our task is to establish the existence of a
maximal hypersurface.  To this accomplish this, we establish the
existence of upper bounds on $a$, $A$, and their inverses on the
interval $[t_0,0)$.  We then introduce a new time function $\tau = f
\circ t$ by introducing a function $f$ that allows us to avoid the
problem associated with $t$ being a bad coordinate on maximal
hypersurfaces.  Once this has been accomplished, theorem~1 will follow
from an argument similar to that used in \cite{Rendall95}.

First, we establish upper bounds on the area radius $r = aA$, the mass
function $m = {1 \over 2} r(1-\nabla^a r\nabla_a r)$, the volume
$V(t)$ of level surfaces of $t$, and their inverses.  That $r$ and
$m^{-1}$ are bounded above follows from the results of
\cite{Burnett91}.  (Note, $m$ is positive.)\ \ Further, the technique
introduced in \cite{MalecOMurchadha94} was used in \cite{Rendall95} to
show that $m/r$ is bounded above on $[t_0,0)$.  Therefore, $m$ and
$r^{-1}$ are also bounded above on $[t_0,0)$.  (That is, the mass $m$
cannot become arbitrarily large and $r$ cannot become arbitrarily
small in this portion of the spacetime.  This is nontrivial as both
$m$ and $r^{-1}$ can become arbitrarily large on unbounded intervals, e.g.,
near an initial or final singularity.)\ \ As we have already
established that the volume of all spatial Cauchy surfaces are bounded
above, $V(t)$ is bounded above.  Using the fact that $\partial_t V(t)$
is positive on $[t_0,0)$, as these hypersurfaces are everywhere
expanding, shows that $V$ is bounded from below by a positive
constant, and hence $V^{-1}$ is bounded above on $[t_0,0)$.

Next, that $a$, $A$, and their inverses are bounded above on $[t_0,0)$
now follows easily from the facts that $r=aA$,
\begin{equation}
V(t) = 4\pi \int a^2 A^3 \;{\rm d} x =
4\pi a^{-1} \int r^3 \;{\rm d} x,
\end{equation}
and our upper bounds for $V$, $r$, and their inverses.

Next, we bound $\alpha'$ using the lapse equation
\begin{equation} \label{lapse}
- A^{-3} (A \alpha')' + ( K_{ab}K^{ab} + R_{ab} n^a n^b)\alpha = 1,
\end{equation}
where $K_{ab}$ is the extrinsic curvature of the CMC hypersurface,
$n^a$ is a unit timelike normal to the CMC hypersurface, and a prime
denotes a derivative by $\partial_x$.  (This is equation~(2.4) in
\cite{Rendall95}.)\ \ Using the fact that $K_{ab}K^{ab}$ is manifestly
non-negative and $R_{ab}n^an^b \ge 0$ by the timelike convergence
condition, it follows that $(A\alpha')' \ge - A^3$.  Using the fact
that $A$ is bounded above and integrating in a CMC hypersurface, we
find that $(A \alpha') |_p - (A \alpha') |_q \ge -C_1$ for some
positive constant $C_1$ and any two points $p$ and $q$ in the
hypersurface.  Choosing $q$ where $\alpha$ is extremal on the surface
(so $\alpha'(q)=0$) and using the fact $A^{-1}$ is bounded above shows
that $\alpha'$ is bounded from below.  Choosing $p$ where $\alpha$ is
extremal on the surface (so $\alpha'(p)=0$) and using the fact
$A^{-1}$ is bounded above shows that $\alpha'$ is bounded from above.
Therefore, there exists a constant $C_2$ such that $|\alpha'| \le
C_2$.  Thus, even if $\alpha$ is unbounded, it must diverge in a way
that is uniform in space: For any two points $p$ and $q$ in a CMC
hypersurface, $|\alpha(p) - \alpha(q)| = |\int_p^q \alpha' \;{\rm d}
x| \le \int_p^q |\alpha'| \;d x \le \pi C_2$.

If we knew that $\alpha$ were bounded above on $[t_0,0)$, we could
then proceed to argue as in \cite{Rendall95}.  While such a bound can
be established rather easily for fields satisfying the dominant energy
and non-negative pressures conditions, such an argument fails for the
massless scalar field.  The difficulty in establishing an upper bound
on $\alpha$ is linked to the possibility that ${\rm d} t$ may be zero
on a maximal hypersurface, and thus $t$ being a bad coordinate.  Note
that this can only occur if $K_{ab}=0$ everywhere on $\Sigma$ (i.e.,
$\Sigma$ is momentarily static) and $R_{ab}n^a n^b = 0$ everywhere on
$\Sigma$.  If the non-negative energy condition ($G_{ab}t^at^b \ge 0$
for all timelike $t^a$) and non-negative sum-pressures condition
[$G_{ab}(t^at^b+g^{ab}) \ge 0$ for all unit-timelike $t^a$] are
satisfied, then $R_{ab}n^a n^b = 0$ implies that $G_{ab}n^an^b =0$
and, hence, by the Hamiltonian constraint equation, the Ricci scalar
curvature of the metric induced on $\Sigma$ must be zero.  However, it
is easy to show that there are no such spherically symmetric
geometries on $S^1 \times S^2$.  Thus, the Einstein-Vlasov spacetimes
do not admit such surfaces.  However, it can be shown that there are
massless scalar field spacetimes with such ``degenerate'' maximal
hypersurfaces.  To avoid this difficulty, we change our time function
to one that is guaranteed to be well-behaved even on a maximal
hypersurface with ${\rm d} t=0$.

Fix any inextendible timelike curve $\gamma$ that is everywhere
orthogonal to the CMC hypersurfaces.  The length of the segment of
$\gamma$ between any two CMC hypersurfaces $t=t_1$ and $t=t_2$ is then
simply $\int_{t_1}^{t_2} \alpha(\gamma(u)) \;{\rm d} u$.  Using the
fact that there is a finite upper bound on the lengths of all timelike
curves in our spacetime, the integral
\begin{equation}
\int_{t_1}^0 \alpha(\gamma(u)) \;{\rm d} u = \lim_{t_2 \to 0}
\int_{t_1}^{t_2} \alpha(\gamma(u)) \;{\rm d} u
\end{equation}
must exist, i.e., $\alpha(\gamma(t))$ is integrable on any interval of
the form $[t_1,0)$. Fix some value $x_0$ of $x$ and consider the
function $\alpha(t,x_0)$. Since $\alpha'$ is bounded there is a
constant $C$ such that $\alpha(t,x_0)\le\alpha(\gamma(t))+C$. It
follows that $\alpha(t,x_0)$ is also integrable on any interval of the
form $[t_1,0)$. Using this fact, define the function $f$ on
$(-\infty,0)$ by setting
\begin{equation} \label{diffeo}
f(\lambda) = \lambda - \int_\lambda^0 \alpha(u,x_0) \;{\rm d} u.
\end{equation}
Noting that $f'(\lambda) = 1 + \alpha(\lambda,x_0)$ and
$\lim_{\lambda \to 0}f(\lambda) = 0$, we see that $f$ is an
orientation-preserving diffeomorphism from $(-\infty,0)$ to
$(-\infty,0)$.  Hence,
\begin{equation} \label{newtime}
\tau = f \circ t
\end{equation}
is a new time function on our spacetime.  Note that
${\partial \tau / \partial t} = 1 + \alpha(t,x_0)$.

The level surfaces of $\tau$ clearly coincide with those of $t$ and so
are CMC hypersurfaces.  As a consequence the field equations for the
geometry and the matter written in terms of $\tau$ look very similar
to those written in terms of $t$.  Using $\tau$ in place of $t$, the
metric has the same form as before
\begin{equation} \label{metric2}
g = -\tilde{\alpha}^2 {\rm d} \tau^2 +
    A^2[({\rm d} x+\tilde{\beta} {\rm d} \tau)^2 + a^2 \Omega],
\end{equation}
where the new lapse function $\tilde{\alpha}$ is given by
\begin{equation} \label{alphat}
\tilde{\alpha} = \alpha \left({\partial t \over \partial \tau}\right)
               = {\alpha \over 1 + \alpha(t,x_0)},
\end{equation}
and similarly for the new shift $\tilde{\beta}$.  In terms of our new
coordinates ($\tau$ replacing $t$) and new variables ($\tilde{\alpha}$
and $\tilde{\beta}$ replacing $\alpha$ and $\beta$, respectively), the
field equations are the same as in \cite{Rendall95} with
$\partial_\tau$ replacing $\partial_t$, $\tilde{\alpha}$ replacing
$\alpha$, $\tilde{\beta}$ replacing $\beta$, and $\partial
t/\partial\tau$ replacing the right-hand side of
equation~(\ref{lapse}).  Explicit occurrences of $t$ in the equations
are left unchanged, $t$ being simply considered as a function of
$\tau$, determined implicitly by equation~(\ref{newtime}).  Using
equation~(\ref{alphat}), it is straightforward to show that $\partial
t / \partial \tau = 1 - \tilde{\alpha}(\tau,x_0)$.  With this, the
lapse equation can be written as
\begin{equation} \label{newlapse}
-A^{-3} (A \tilde\alpha')' + ( K_{ab}K^{ab} + R_{ab} n^a n^b)\tilde\alpha
= 1 - \tilde\alpha(\tau,x_0).
\end{equation}
Using the fact that $\alpha'$ is bounded, as argued above, it follows
that $\alpha (t,x)\le \alpha(t,x_0) + C$, where $C$ is a constant.
Therefore, by equation~(\ref{alphat}), $\tilde{\alpha}$ is bounded
above.

It is now possible to apply the same type of arguments to the system
corresponding to the time coordinate $\tau$ as were applied in
\cite{Rendall95} to the system corresponding to the time coordinate
$t$ to show that all the basic geometric and matter quantities in the
equations written with respect to $\tau$ are bounded and that the same
is true for their spatial derivatives of any order.  Bounding time
derivatives of all these quantities requires some more effort.  All
but one of the steps in the inductive argument used to bound time
derivatives in \cite{Rendall95} apply without change.  (Note that in
\cite{Rendall95}, derivatives with respect to $t$ were bounded,
whereas here, derivatives with respect to $\tau$ are bounded.)\ \ The
argument that does not carry over is that which was used to bound time
derivatives of $\alpha$ and $\alpha'$.  To see why, consider the
equation obtained by differentiating equation~(\ref{newlapse}) $k$
times with respect to $\tau$
\begin{eqnarray} \label{derivatives}
 - A^{-3} (A (D^k_\tau \tilde\alpha)')'
 +  ( K_{ab}K^{ab} + R_{ab} n^a n^b) D^k_\tau \tilde\alpha\nonumber\\
 + D^k_\tau \tilde\alpha(\tau,x_0) = B_k,
\end{eqnarray}
where $D^k_\tau = \partial^k_\tau$ denotes the $k$-th partial
derivative with respect to $\tau$.  Here $B_k$ is an expression
which is already known to be bounded when we are at the step in the
inductive argument to bound $D^k_\tau\tilde\alpha$ and
$D^k_\tau\tilde\alpha'$.  In lemma~3.4 of \cite{Rendall95}, $D^k_t
\alpha$ was bounded by using the fact that $t$ was bounded away from
zero.  The analogous procedure is clearly not possible in the present
situation, where $t$ is tending to zero.  This kind of argument was
also used in \cite{Rendall95} to bound time derivatives of higher
order spatial derivatives of $\alpha$, but that is unnecessary, since
such bounds can be obtained directly by differentiating the lapse
equation once the time derivatives of $\alpha$ and $\alpha'$ have been
bounded.  The same argument applies here, so all we need to do is to
prove the boundedness of $D^k_\tau\tilde\alpha$ and
$D^k_\tau\tilde\alpha'$ using equation~(\ref{derivatives}) under the
hypothesis that $B_k$ is bounded.  This follows by simply noting that
equation~(\ref{derivatives}) has the same form for each value of $k$
and the following lemma.

{\it Lemma 1.} Consider the differential equation

\begin{equation} \label{ODE}
(au')'=bu+c+du(x_0)
\end{equation}
where $a$, $b$, $c$, $d$, and $u$ are $2\pi$-periodic functions on the
real line and $x_0$ is a point therein.  Suppose that $a>0$, $b\ge 0$,
$d\ge 0$, and that $d$ is not identically zero.  Then $|u|$ and $|u'|$
are bounded by constants depending only on the quantities $K_1 =
\max\{a^{-1}(x)\} > 0$, $K_2=\int_0^{2\pi} |c(x)| \;{\rm d} x \ge 0$,
$K_3 = \int_0^{2\pi} d(x)\;{\rm d} x > 0$, and $K_4 = \int_0^{2\pi}
b(x) \;{\rm d} x \ge 0$.

{\it Proof.}  First, if $u(x_0) > 2\pi K_1K_2$, then $u > 0$
everywhere.  To see this, suppose otherwise and let $x_1$ be a point
where $u$ achieves its maximum, so $u(x_1) \ge u(x_0) > 2\pi K_1K_2$
and let $x_2$ be that number such that $u > 0$ on $[x_1,x_2)$ and
$u(x_2)=0$ (so $x_1 < x_2 < x_1 + 2\pi$).  Then, on the interval
$[x_1,x_2]$, we have $(au')' \ge c$, from which it follows that $u'
\ge -K_1K_2$ on $[x_1,x_2]$.  Integrating this and using the fact that
$u(x_2)=0$, we find that $u(x_1) \le 2\pi K_1K_2$, contradicting the
fact that $u(x_1) > 2\pi K_1K_2$.  Therefore, as $u$ is everywhere
positive, it follows that $(au')'\ge c$. Integrating this inequality
starting (or ending) at a point where $u' = 0$ shows that $|u'| \le
K_1K_2$.  Integrating equation~(\ref{ODE}) from $0$ to $2\pi$ and
using the fact that $u$ is positive gives $u(x_0)\int_0^{2\pi} d(x)
\;{\rm d} x \le \int_0^{2\pi} |c(x)| \;{\rm d} x$, and hence,
$|u(x_0)| \le K_2 K_3^{-1}$.  Using this and the fact that $|u'| \le
K_1K_2$ shows that $|u| \le K_2 K_3^{-1} + 2\pi K_1K_2$.  Second, if
$u(x_0) < -2\pi K_1K_2$, a similar argument shows that $u$ is
everywhere negative and we again obtain the same bounds on $|u'|$ and
$|u|$.  Third, suppose that $|u(x_0)| \le 2\pi K_1K_2$.  If $\max(u) >
2\pi K_1K_2(1 + 2\pi K_1K_3)$, using the inequality $(au')' \ge c +d
u(x_0)$, we can argue much as before to see that $u$ is everywhere
positive and again obtain the same bounds on $|u'|$ and $|u|$.
Similarly, if $\min(u) < - 2\pi K_1K_2(1 + 2\pi K_1K_3)$, it follows
that $u$ is everywhere negative and again we recover the same bounds
on $|u'|$ and $|u|$.  Next, if $|u|
\le 2\pi K_1K_2(1 + 2\pi K_1K_3)$ everywhere, $|u|$ is already
bounded, and to bound $|u'|$, we note that we have bounds for all
terms on the right hand side of equation~(\ref{ODE}), so it suffices
to integrate it starting from a point where $u'$ is zero to bound
$|u'|$.$\Box$

At this stage, we have indicated how all geometric and matter
quantities, expressed in terms of the new time coordinate $\tau$,
can be bounded, together with all their derivatives. In particular,
this means that all these quantities are uniformly continuous
on any interval of the form $[\tau_1,0)$, where $\tau_1$
is finite. It follows that all these quantities have smooth extensions
to the interval $[\tau_1,0]$. Restricting them to the hypersurface
$\tau=0$ gives a initial data set for the Einstein-matter
equations with zero mean curvature. By the standard uniqueness
theorems for the Cauchy problem, the spacetime which, in the old
coordinates, was defined on the interval $(-\infty,0)$ is isometric
to a subset of the maximal development of this new initial data
set. It follows that the original spacetime has an extension which
contains a maximal hypersurface.

Lastly, that the foliation is unique now follows from the fact that
compact CMC Cauchy surfaces with non-zero mean curvature are unique
\cite{MarsdenTipler80} and that the spacetime is indeed maximal
follows from the fact that any spacetime admitting a complete foliation
by compact CMC Cauchy surfaces is maximal \cite{EardleySmarr79}.

\section{A bound for the volume of space} \label{sec:volume}

It is well known that as we transport an ``infinitesimal'' spacelike
surface $S$ along the geodesics normal to itself, the ratio $\nu$ of
its volume of to its original volume is governed by the Raychaudhuri
equation
\begin{equation} \label{ray}
{{\rm d}^2 \over {\rm d} t^2} \nu^{1/3} + {1 \over 3}
\left( R_{ab}t^at^b + \sigma_{ab}\sigma^{ab} \right) \nu^{1/3} = 0,
\end{equation}
where $t$ is the proper time measured along the geodesics normal to
$S$, $R_{ab}$ is the Ricci tensor, and $\sigma_{ab}$ is the shear
tensor associated with the geodesic flow
\cite{HawkingEllis73,Wald84,Penrose72}.  (This equation is usually
written in terms of the divergence of the geodesic flow $\theta =
\nu^{-1} d\nu/dt$.)\ \ On the surface $S$, $\nu$ satisfies the initial
condition $\nu = 1$ and $d\nu/dt = - H(p)$, where $H(p)$ is the trace
of the extrinsic curvature of $S$ at the point $p$ where the geodesic
intersects $S$.  Therefore, if the spacetime satisfies the
timelike-convergence condition ($R_{ab}t^at^b \ge 0$ for all timelike
$t^a$), it follows that as long as $\nu$ remains non-negative,
\begin{equation}
{{\rm d}^2 \over {\rm d} t^2} \nu^{1/3} \le 0,
\end{equation}
from which we find that
\begin{equation} \label{localbound}
\nu(t) \le \left[1 - {1 \over 3} H(p) (t-t_0) \right]^3.
\end{equation}
This equation bounds the growth of the volume of a local spatial
region in the spacetime.

Using this result, it is not difficult to show that, in a spacetime
satisfying the timelike-convergence condition, if we fix a Cauchy
surface $\Sigma_0$ and construct from it a second Cauchy surface
$\Sigma$ by transporting $\Sigma_0$ to the future along the flow
determined by the geodesics normal to $\Sigma_0$, as long as these
flow lines do not self-intersect (which will be true if $\Sigma$ is
sufficiently close to $\Sigma_0$), then
\begin{equation} \label{specialglobalbound}
\text{vol}(\Sigma) \le \text{vol}(\Sigma_0) \left[ 1 + {1 \over 3}
\sup_{\Sigma_0}(-H) T \right]^3,
\end{equation}
where $\text{vol}(S)$ denotes the three-volume of a Cauchy surface $S$
and $T$ is the ``distance'' between the two surfaces measured by the
lengths of the geodesics normal to $\Sigma_0$ (which will be
independent of which geodesic is chosen by the construction of
$\Sigma$).  Therefore, we have a bound on the volume of $\Sigma$ in
terms of the volume of $\Sigma_0$, the extrinsic curvature of
$\Sigma_0$, and the distance between $\Sigma_0$ and $\Sigma$.  Does a
similar result hold for more general Cauchy surfaces $\Sigma$?  For
instance, a more general hypersurface $\Sigma$ may not be everywhere
normal to the geodesics from $\Sigma_0$, some geodesics normal to
$\Sigma_0$ may intersect one another between $\Sigma_0$ and $\Sigma$,
and parts of $\Sigma$ may lie to the future of $\Sigma_0$ while other
parts may lie to the past.  Can the simple bound given by
equation~(\ref{specialglobalbound}) be modified to cover these
cases?  That it can is the subject of the following lemma.

{\it Lemma 2.} Fix an orientable globally hyperbolic spacetime
$(M,g_{ab})$ satisfying the timelike-convergence condition
($R_{ab}t^at^b \ge 0$ for all timelike $t^a$) and a smooth spacelike
Cauchy surface $\Sigma_0$ therein.  Then, for any smooth spacelike
Cauchy surface $\Sigma$,
\begin{equation} \label{boundany}
\text{vol}(\Sigma) \le \text{vol}(\Sigma_0)
\left[ 1 + {1 \over 3} \sup_{\Sigma_0}(|H|)
\Delta(\Sigma_0,\Sigma) \right]^3,
\end{equation}
where $\text{vol}(S)$ denotes the three-volume of a Cauchy surface
$S$, $H$ is the trace of the extrinsic curvature of $\Sigma_0$ (using
the convention that $H$ measures the {\it convergence} of the {\it
future-directed} timelike normals to a spacelike surface), and
$\Delta(\Sigma_0,\Sigma)$ is the least upper bound to the lengths of
causal curves connecting $\Sigma_0$ to $\Sigma$ (either future or past
directed).  Further, for any Cauchy surface $\Sigma \subset
D^+(\Sigma_0)$,
\begin{equation} \label{boundfuture}
\text{vol}(\Sigma) \le \text{vol}(\Sigma_0)
\left[ 1 + {1 \over 3} \sup_{\Sigma_0}(-H)
\Delta(\Sigma_0,\Sigma) \right]^3.
\end{equation}

Note that for $p,q \in M$, $\Delta(p,q)$ is not quite the distance
function $d(p,q)$ as used in \cite{HawkingEllis73} as $d(p,q) =
0$ if $q \in J^-(p)$.  Instead, $\Delta(p,q)$ does not distinguish
between future and past: $\Delta(p,q) = \Delta(q,p) = d(p,q) +
d(q,p)$.

{}From lemma~2, we see that for a spacetime satisfying the
timelike-convergence condition, possessing compact Cauchy surfaces,
and having a finite lifetime (in the sense that $d(p,q)$ [equivalently
$\Delta(p,q)$] is bounded above by a constant independent of $p$ and
$q$), then the volume of a Cauchy surface therein cannot be
arbitrarily large.  Further, we see that if the spacetime admits a
maximal Cauchy surface $\Sigma_0$ ($H=0$ thereon), we reproduce the
result that there is no other Cauchy surface having volume larger than
$\Sigma_0$ (though there may be surfaces of equal volume)
\cite{MarsdenTipler80}.

In the following, ${\rm d} f$ denotes the derivative map associated
with a differentiable map $f$ between manifolds.  When viewed as a
pull-back, we denote ${\rm d} f$ by $f^*$ and, when viewed as a
push-forward, we denote ${\rm d} f$ by $f_*$.  For a map $f: A \to B$,
$f[A]$ denotes the image of $A$ in $B$.  Lastly, $A \setminus B$
denotes the set of elements in $A$ that are not in $B$.

\subsection{Proof of lemma 2} \label{sec:prooflemma2}

To begin the proof of lemma~2, for each point $p \in \Sigma_0$, let
$\gamma_p$ denote the unique inextendible geodesic containing $p$ and
intersecting $\Sigma_0$ orthogonally.  Parameterize $\gamma_p$ by $t$
so that the tangent vector to $\gamma_p$ is future-directed
unit-timelike and $\gamma_p(0)=p$.  Then, define the map $f:\Sigma_0
\to \Sigma$, by
\begin{equation}
f(p) = \gamma_p \cap \Sigma.
\end{equation}
Note that for each $p \in \Sigma_0$, $f$ is well defined since
$\gamma_p$ intersects $\Sigma$ at precisely one point as $\Sigma$
is a spacelike Cauchy surface for the spacetime.

Next, let ${\cal K}$ be the subset of $\Sigma_0$ defined by the
property that $p \in {\cal K}$ if and only if the geodesic $\gamma_p$
does not possess a point conjugate to $\Sigma_0$ between $\Sigma_0$
and $\Sigma$ (although it may have such a conjugate point on
$\Sigma$).  Note that this is precisely the condition that for each $p
\in {\cal K}$ the solution $\nu$ to equation~(\ref{ray}) along
$\gamma_p$, satisfying the initial conditions $\nu = 1$ and $d\nu/dt =
H(p)$ at $p$, be strictly positive on the portion of $\gamma_p$
between $p$ and $f(p)$.  It follows that ${\cal K}$ is closed.
Furthermore, $f$ maps ${\cal K}$ onto $\Sigma$.  To see this, recall
that for any point $q \in \Sigma$ there exists a timelike curve $\mu$
connecting $q$ to $\Sigma_0$ having a length no less than any other
such curve.  Furthermore, such a curve $\mu$ must intersect $\Sigma_0$
normally, is geodetic, and has no point conjugate to $\Sigma_0$
between $\Sigma_0$ and $q$.  (See Theorem~9.3.5 of \cite{Wald84}.)\ \
Therefore, the point $p = \mu \cap \Sigma_0$ is in ${\cal K}$ and $\mu
\subset \gamma_p$, so $f(p) = \gamma_p \cap \Sigma = \mu \cap \Sigma =
q$.  Therefore, $f$ maps ${\cal K}$ onto $\Sigma$.  However, in
general, $f$ will not be one-to-one between ${\cal K}$ and $\Sigma$.

Let $C$ denote the set of critical points of the map $f$ on
$\Sigma_0$.  That is, $p \in C$ if and only if its derivative map
$f_*: (T\Sigma_0)_p \to (T\Sigma)_{f(p)}$ is not onto.  Then, by
Sard's theorem \cite{Milnor65}, $f[C]$ (the critical values of $f$),
and hence $f[{\cal K} \cap C]$, are sets of measure zero on $\Sigma$.
Now, note that $\Sigma$ can be expressed as the union of $f[{\cal K}
\setminus C]$ and a set having measure zero.  To see this, we write
\begin{eqnarray}
\Sigma & = & f[{\cal K}]
 = f[({\cal K} \setminus C) \cup ({\cal K} \cap C)] \nonumber\\
& = & f[{\cal K} \setminus C] \cup
\left( f[{\cal K} \cap C] \setminus f[{\cal K} \setminus C] \right).
\end{eqnarray}
The last two sets are manifestly disjoint and the latter is a set of
measure zero (as it is a subset of a set of measure zero).  Therefore, we
need only concern ourselves the behavior of $f$ on the set of regular
points of $f$ within ${\cal K}$.  This is useful since, by the inverse
function theorem \cite{Milnor65}, $f$ is a local diffeomorphism
between ${\cal K} \setminus C$ and $f[{\cal K} \setminus C]$.  As we
shall see, for all $p \in {\cal K}\setminus C$, the point $f(p)$ is
not conjugate to $\Sigma_0$ on $\gamma_p$, from which it follows that
${\cal K} \setminus C$ is an open subset of $\Sigma_0$.

Denote volume elements associated with the induced metrics on
$\Sigma_0$ and $\Sigma$ by $e_{abc}$ and $\epsilon_{abc}$,
respectively, chosen so that $e_{abc}$ and $\epsilon_{abc}$ correspond
to the same spatial orientation class (which can be done as the
spacetime is both time-orientable and orientable).  Then the Jacobian
of the map $f$ is that unique scalar field $J$ on $\Sigma_0$ such that
\begin{equation} \label{defJ}
(f^*\epsilon)_{abc} = J e_{abc}.
\end{equation}
Note that $J$ is zero on $C$ and positive on ${\cal K} \setminus C$.

With these definitions, we have
\begin{eqnarray} \label{boundvol}
\text{vol}(\Sigma)
& = & \int_{f[{\cal K} \setminus C]} \epsilon \nonumber \\
& \le & \int_{{\cal K} \setminus C} (f^*\epsilon) \nonumber \\
& \le & \left[\sup_{{\cal K} \setminus C}(J)\right]
\int_{{\cal K} \setminus C} e \nonumber \\
& \le &   \left[\sup_{{\cal K} \setminus C}(J)\right]
\text{vol}(\Sigma_0).
\end{eqnarray}
The first step follows from the facts that $\Sigma = f[{\cal K}]$ and
$f[{\cal K} \cap C]$ is a set of measure zero.  That we have an
inequality in the second step follows from the fact that although $f$
is a local diffeomorphism, it may not be one-to-one between ${\cal K}
\setminus C$ and $f[{\cal K} \setminus C]$.  The third step follows
from the definition of $J$ given by equation~(\ref{defJ}) and the fact
that $J$ is bounded above by its supremum.  Lastly, the fourth step
follows from the fact that ${\cal K} \setminus C$ is a subset of
$\Sigma_0$.  So, to prove lemma~2, we need to show that, on the set
${\cal K} \setminus C$, $J$ is bounded above by the relevant
expressions in lemma~2.

To that end, define $\phi: \Sigma_0 \times {\Bbb R} \to M$ by setting
$\phi(p,t) = \gamma_p(t)$.  Of course, if $\gamma_p$ is not future and
past complete, this will not be defined for all $t$.  Next, define
$T:\Sigma_0 \to {\Bbb R}$ by setting $T(p)$ to that number such that
$\gamma_p(T(p)) = f(p)$, i.e., $T(p)$ is the ``time'' along the
geodesic $\gamma_p$ at which $\gamma_p$ intersects $\Sigma$.  Note
that if $f(p)$ lies to the future of $\Sigma_0$, then $T(p)$ is
positive, while if $f(p)$ lies to the past of $\Sigma_0$, then $T(p)$
is negative.

Fix a point $p \in {\cal K} \setminus C$ and define the map $g:
\Sigma_0 \to M$ by setting $g(q) = \phi(q,T(p))$.  Should
$\gamma_q(T(p))$ not be defined, then $g$ is not defined for that
point of $\Sigma_0$.  However, it will always be defined for some
neighborhood of $p$ as $g(p) = f(p)$.  Notice that $g$ simply
``translates'' points on $\Sigma_0$ along the geodesics normal to
$\Sigma_0$ a fixed distance $T(p)$ (independent of point), i.e., it is
a translation along the normal geodesic ``flow''.  Therefore, the
derivative map of $g$ at a point is precisely the geodesic deviation
map.  In particular, ${\rm d} g$ is injective (one-to-one) from
$(T\Sigma_0)_p$ to $(TM)_{f(p)}$ if and only if $f(p)$ is not
conjugate to $\Sigma_0$ on $\gamma_p$ (by the definition of such a
conjugate point).

Noting that $f$ can be written as $f(q) = \phi(q,T(q))$, we see that
the derivative maps of $f$ and $g$ at $p$ [both of which are maps from
$(T\Sigma_0)_p$ to $(TM)_{f(p)}$] are related by
\begin{equation} \label{dmaps}
({\rm d} f)^a{}_b = ({\rm d} g)^a{}_b + t^a ({\rm d} T)_b,
\end{equation}
where $t^a$ is the unit future-directed tangent vector to $\gamma_p$
at $f(p)$.  From this we see that ${\rm d} f$ is injective [from
$(T\Sigma_0)_p$ to $(TM)_{f(p)}$] if and only if ${\rm d} g$ is
injective.  Therefore, on ${\cal K} \setminus C$, not only is ${\rm d}
f$ injective, but ${\rm d} g$ is also injective, and hence $f(p)$ is
not conjugate to $\Sigma_0$ on $\gamma_p$.

Define $\hat{e}_{abc}$ at $f(p)$ by parallel transporting $e_{abc}$ at
$p$ along $\gamma_p$.  Then,
\begin{equation} \label{f^*e-hat}
(f^*\hat{e})_{abc} = (g^*\hat{e})_{abc} = \nu(T(p)) e_{abc}.
\end{equation}
The first equality follows from (\ref{dmaps}) and the fact that $t^a
\hat{e}_{abc} = 0$. The second equality follows by recognizing that
the coefficient of the right-hand most term is precisely the ratio of
the volume of an ``infinitesimal'' region in $\Sigma_0$ to its
original volume as it is transported along the geodesic flow normal to
$\Sigma_0$.  As the transport is done from $p$ to $f(p)$, the
coefficient is $\nu(T(p))$, where $\nu$ is the solution of
equation~(\ref{ray}) satisfying the stated initial conditions.  (In
other words, $\nu(t)$ is the Jacobian of the geodesic deviation map.)

Denote the future-directed normal to $\Sigma$ at $f(p)$ by $n^a$.
Then, there exists a unit-spacelike vector $x^a \in (T\Sigma)_{f(p)}$
such that $t^a = \gamma (n^a + \beta x^a)$, where $\gamma = (-t^a
n_a)$ and $\beta = \sqrt{1 - \gamma^{-2}}$.  Then, for one of the two
volume elements $\epsilon_{abcd}$ on $M$ associated with the spacetime
metric, we have $\epsilon_{abc} = n^m \epsilon_{mabc}$ and
$\hat{e}_{abc} = t^m \epsilon_{mabc}$, which gives the following
relation between these two tensors at $f(p)$,
\begin{equation} \label{relate}
\hat{e}_{abc} = \gamma \epsilon_{abc} + \gamma\beta x^m\epsilon_{mabc}.
\end{equation}
Therefore,
\begin{equation} \label{relatepullbacke-hatepsilon}
(f^* \hat{e})_{abc} = \gamma (f^*\epsilon)_{abc},
\end{equation}
where we have used (\ref{relate}) and the fact that the pull-back of
$x^m \epsilon_{mabc}$ by $f$ must be zero as $x^m$ is in the surface
$\Sigma$ and the contraction of $\epsilon_{abcd}$ with four vectors
all in a three-dimensional subspace must be zero.  Therefore, using
(\ref{relatepullbacke-hatepsilon}) and~(\ref{f^*e-hat}), we see that
\begin{equation} \label{f^*epsilon}
(f^*\epsilon)_{abc} = (-t^an_a)^{-1}\nu(T(p)) e_{abc},
\end{equation}
which when compared to (\ref{defJ}), gives
\begin{equation}
J(p) = (-t^an_a)^{-1}\nu(T(p)).
\end{equation}

Since $(-t^a n_a)^{-1} \le 1$ and $\nu(T(p))$ is bounded above by
(\ref{localbound}), we have
\begin{equation}
J(p) \le \left[ 1 - {1 \over 3} H(p)T(p) \right]^3.
\end{equation}
So, if $\Sigma \subset D^+(\Sigma_0)$, we have $0 \le T(p) \le
\Delta(\Sigma_0,\Sigma)$ and $-H(p) \le \sup_{\Sigma_0}(-H)$, and
therefore,
\begin{equation}
\sup_{{\cal K} \setminus C}(J) \le \left[1 + {1 \over 3}
\sup_{\Sigma_0}(-H)\Delta(\Sigma_0,\Sigma) \right]^3,
\end{equation}
which with (\ref{boundvol}) establishes
equation~(\ref{boundfuture}).  More generally, as
\begin{equation}
H(p)T(p) \le |H(p)||T(p)| \le \sup_{\Sigma_0}(|H|)
\Delta(\Sigma_0,\Sigma),
\end{equation}
we have
\begin{equation}
\sup_{{\cal K} \setminus C}(J) \le \left[1 + {1 \over 3}
\sup_{\Sigma_0}(|H|)\Delta(\Sigma_0,\Sigma) \right]^3,
\end{equation}
which with (\ref{boundvol}) establishes equation~(\ref{boundany}).
This completes the proof of lemma~2.

\end{document}